\documentclass[journal ]{new-aiaa}
\usepackage[utf8]{inputenc}
\usepackage{textcomp}

\usepackage{mathtools}
\usepackage{graphicx}

\usepackage{amsmath}
\usepackage[version=4]{mhchem}
\usepackage{siunitx}
\usepackage{longtable,tabularx}
\usepackage{soul}
\usepackage{ulem}
\usepackage{comment}
\usepackage{subcaption}
\usepackage{diagbox}
\usepackage{xcolor}
\usepackage{soul}
\usepackage{color}
\definecolor{mycolor1}{rgb}{1, 0, 0}

\definecolor{mycolor2}{rgb}{0, 0, 1}

\definecolor{mycolor3}{rgb}{0, 0.5, 0.5}

\usepackage{humberto_commands}
\usepackage{humberto_variable_names}

\title{A summary on the UD Kalman Filter}

\author{J. Humberto Ramos\footnote{Postdoctoral Researcher, Mechanical and Aerospace Engineering, AIAA Member, jramoszuniga@ufl.edu}}
\affil{University of Florida, Shalimar, FL 32579 }
\author{Kevin M. Brink\footnote{Senior Research Engineer, Senior Member AIAA.}}
\affil{U.S. Air Force Research Laboratory, Eglin Air Force Base, FL 32542}
\author{Prashant Ganesh\footnote{Research Assistant Engineer, Mechanical and Aerospace Engineering}}
\affil{University of Florida, Shalimar, FL}
\author{John E. Hurtado\footnote{Professor, Department of Aerospace Engineering and Interim Dean of Engineering, College of Engineering.}}
\affil{Texas A\&M University, College Station, TX 77843}

\begin{document}
	
	\maketitle
	
	\begin{abstract}
	    This document contains a concise and unified reference for one of the existing mechanizations of the UD Kalman filter. The associated matrix algorithms are also included along with the corresponding references.
	\end{abstract}
	
\section{Introduction}\label{sec:background}
The UD formulation of the Kalman filter does not operate on the covariance matrix $\Cov$ directly. Instead, it updates and propagates the covariance factors $\Umat$ and $\D$ ($\Cov=\Umat\D\Umat\trans$). This formulation is preferred for hardware implementation as it provides an efficient and numerically stable solution while ensuring the symmetry of the covariance matrix even for large filters \cite{Thornton1976}. In contrast with Potter's square root filter \cite{Kaminski1971} that uses Cholesky factors $\Smat$, such that $\Cov = \Smat\Smat\trans $, the UD formulation does not necessarily increase the numerical precision of the filter \cite{d2018information}, but it is generally preferred for its relatively high efficiency. 

The following sections summarize the UD Kalman filter, and present the details on its implementation. This independent document is also intended to be a companion paper to a more general UD filter presented in \cite{UD_partial}.

\section{The UD Factorization in Kalman Filtering}\label{sec:ud_factorization_intro_propagation} The UD factors can be obtained as a corollary of the Cholesky square-root factors $\Smat$ used in Potter's filter (\cite{bierman2006factorization}, Chap. 10). A closer examination of the Cholesky decomposition of an $n\times n$ matrix, reveals that $n$ square root operations are computed and that those same square-roots appear dividing each column of the matrix, motivating a pair of alternative Cholesky factors: the UD factors. To illustrate this and further see how the UD factors can be obtained, consider a $ 3 \times 3 $ covariance matrix, $ \Cov $, to be analytically factorized via Cholesky decomposition as
\begin{equation}
	\Cov = \Smat\Smat\trans \ 
\end{equation}
with $ \Smat $ written as
%

\begin{equation}
	\Smat = \begin{bmatrix}
		\sqrt{\Cov_{11}- \frac{\Cov_{13}^2}{\Cov_{33}}-\frac{(\Cov_{12}-\frac{\Cov_{13}\Cov_{23}}{\Cov_{33}})^2}{\Cov_{22}-\frac{\Cov_{23}^2}{\Cov_{33}}}} &
		\frac{\Cov_{12}-\frac{\Cov_{13}\Cov_{23}}{\Cov_{33}}} {\sqrt{\Cov_{22}-\frac{\Cov_{23}^2}{\Cov_{33}}}}& \frac{\Cov_{13}}{{\sqrt{\Cov_{33}}}}\\
		0 		&	\sqrt{\Cov_{22}-\frac{\Cov_{23}^2}{\Cov_{33}}}	& \frac{\Cov_{23}}{{\sqrt{\Cov_{33}}}}\\
		0 & 0 & \sqrt{\Cov_{33}}
	\end{bmatrix} \ 
\end{equation}
which can further be factorized as

\begin{equation}
	\Smat	= \begin{bmatrix}
		1 & \frac{\Cov_{12}-\frac{\Cov_{13}\Cov_{23}}{\Cov_{33}}} {\Cov_{22}-\frac{\Cov_{23}^2}{\Cov_{33}}} & \frac{\Cov_{13}}{\Cov_{33}} \\
		0 & 1 & \frac{\Cov_{23}}{\Cov_{33}} \\
		0 & 0 & 1
	\end{bmatrix}
	\begin{bmatrix}
		\sqrt{\Cov_{11}- \frac{\Cov_{13}^2}{\Cov_{33}}-\frac{(\Cov_{12}-\frac{\Cov_{13}\Cov_{23}}{\Cov_{33}})^2}{\Cov_{22}-\frac{\Cov_{23}^2}{\Cov_{33}}}} & 0 & 0 \\
		0 & \sqrt{\Cov_{22}-\frac{\Cov_{23}^2}{\Cov_{33}}} & 0 \\
		0  & 0 & \sqrt{\Cov_{33}}
	\end{bmatrix} \ 
\end{equation}
Defining the upper triangular matrix with ones on the diagonal as $ \Umat $, and the diagonal matrix as $\sqrt{\D} $, the square root factor $\Smat $ can now be written as
\begin{equation}
	\Smat = \Umat\sqrt{\D} \ 
\end{equation}
which provides the alternative covariance matrix factorization:
\begin{equation}
	\Cov=	 \Smat\Smat\trans = \Umat\sqrt{\D} \sqrt{\D}\Umat\trans = \Umat \D\Umat\trans \ 
\end{equation}

Because the square root terms are contained in the full representation of $\D$, and the filter propagates and updates the matrices $\Umat$ and $\D$, no square root operations are executed in the UD Kalman filter. Moreover, when a Kalman filter is in UD form, monitoring the covariance positive semi-definiteness only requires checking the sign of the elements of $ \D $, and enforcement of its positive semi-definiteness can be done by ensuring they are non-negative \cite{d2018information}. 
A mechanization to obtain the UD factors is included here in Appendix \ref{ap:udu_decomposition}. But many other sources are also available in the filtering literature (\cite{bierman2006factorization}, Chap. 8), (\cite{simon2006optimal}, p. 174), (\cite{Grewal2001}, Chap. 7), in linear algebra books, and in matrix operations books (\cite{strang1993introduction}, p. 563), (\cite{golub2012matrix}, Chap. 5). Further elaboration on this topic, and other factorized variants of Kalman filter can be found in \cite{Gibbs2011a}, \cite{Ramo1907:Square}.

\subsection{UD Kalman Filter Propagation step}\label{subsec:overview_UD} 
There are different alternatives to perform a time update using the UD filter. This document presents the formulation from (\cite{simon2006optimal}, p. 176) and \cite{Carpenter2018} as this form is common in the literature, and it does not assume a particular structure of the system state-space. Alternative UD factor propagation approaches can be found in \cite{Gibbs2011a}.

To illustrate the idea behind the propagation of the UD factors within the Extended Kalman filter (EKF) framework (\cite{crassidis2011optimal}, Chap. 5), consider the following discrete nonlinear dynamic system with state vector $\x_k \in \Real^n$, known input $\uk \in \Real^r$, measurement vector  $\yktilde \in \Real^m$, process noise $ \w_k \in \Real^q$, and measurement noise $\v_k \in \Real^m$:
\begin{equation}\label{eq:nonlinear_equations}
	\x_{k}=\f_{k-1}(\x_{k-1},\u_{k-1},\w_{k-1})
\end{equation}
\begin{equation}
	\yktilde=\hkofxk+\vk
\end{equation}
\begin{equation}
	\wk \sim \mathcal{N} (\zerovec,\Qmat_k)
\end{equation}
\begin{equation}\label{eq:measurement_cov}
	\vk \sim \mathcal{N} (\zerovec,\R_k)
\end{equation}
Here, $\wk$ and $\vk $ are zero-mean Gaussian white-noise processes, with covariances $\Qmat_k=\expect[\wk\wk\trans]$ and $\R_k=\expect[\vk\vk\trans]$, respectively; the function $\hkofxk$ is the measurement model function, and all of the sub-indices denote the time instance. The UD EKF and the conventional EKF propagate states similarly via a nonlinear or linearized process model. Finally, consider the EKF discrete covariance propagation equation given as

\begin{equation}\label{eq:covariance_propagation}
	\stdvec{P}[][k][-] = \Fmat_{k-1}\stdvec{P}[][k-1][+]\Fmat_{k-1}\trans+\Gmat_{k-1}\Qmat_{k-1}\Gmat_{k-1}\trans 
\end{equation}
Here, $\stdvec{P}_k^- $ is the prior error covariance, $\stdvec{P}[][k-1][+]$ is the most recent updated covariance, $ \Gmat_{k-1} = \partder{\f_{k-1}}{\w}
$ is the $ n \times q $ matrix mapping the process noise from the vector $ \wk$ to the state, and $\Fmat_{k-1} = \partder{\f_{k-1}}{\x}$ is the Jacobian of the process. Both, $\Fmat_{k-1}$ and $\Gmat_{k-1}$, are evaluated at $\expect[\x_{k-1}],\ \expect[\w_{k-1}]=0,$ and $ \u_{k-1}$.

The idea of the UD propagation stage is to operate on a factorized form of $\stdvec{P}[][k][-] = \Fmat_{k-1}\stdvec{P}[][k-1][+]\Fmat_{k-1}\trans+\Gmat_{k-1}\Qmat_{k-1}\Gmat_{k-1}\trans $ composed by three factors that resemble the form $ \Cov^-_k = \Usub_k\Dsub_k\Usub_k\trans$ (where the upper bar in a quantity, i.e., $ \bar{[\,\cdot\,]}$, indicates a prior quantity). A direct attempt to factorize $\stdvec{P}[][k][-] = \Fmat_{k-1}\stdvec{P}[][k-1][+]\Fmat_{k-1}\trans+\Gmat_{k-1}\Qmat_{k-1}\Gmat_{k-1}\trans $ into the three factors, $\Wmat\Dhat\Wmat\trans$, leads to the candidate form of (with no time indices for convenience)
\begin{equation}\label{eq:P_wrong_dimensions}
	\stdvec{P}^- = \Usub\Dsub\Usub\trans= \rowvec{\Fmat\Uplus,\Gmat}\begin{bmatrix}
		\Dplus & \zerovec \\ \zerovec & \Qmat
	\end{bmatrix}\colvec{\Uplus\trans\Fmat\trans,\Gmat\trans}= \Wmat\Dhat\Wmat\trans \ 
\end{equation}
where the notation $ \overset{+}{[\,\cdot\,]}$ indicates posterior quantities; in this case, the most recent posterior quantities. At this point 
the propagated $\Usub$ and $\Dsub$ factors from Eq. (\ref{eq:P_wrong_dimensions}) cannot be obtained via a direct term-by-term comparison (i.e. $\Usub\neq\Wmat$, $\Dsub\neq\Dhat$) because the diagonal matrix $ \Dhat $ defined above is an $ (n+q)\times(n+q) $ matrix, and $ \Wmat = [\Fmat\Uplus \ \Gmat] $ is an $ n\times (n+q)$ matrix and is not upper triangular in general. Nevertheless, some work can be done on $\Wmat$ and $\Dhat$ to rewrite them in the appropriate form and dimension such that the propagated $\Usub$ and $\Dsub$ can be obtained via a direct comparison. That is, one seeks to have
\begin{equation}\label{eq:udu_equal_wdw}
	\Cov^-=\Usub{\Dsub}\Usub\trans = \tilde{\Wmat}\tilde{\D}\tilde{\Wmat}\trans  \ \end{equation}
such that $ \Usub=\tilde{\Wmat} $ and $ \Dsub=\tilde{\D}$, where $\tilde{\Wmat}$ and $\tilde{\D}$ are alternative factors to be obtained. In an abuse of notation, the tilde on the $\Wmat$ and $\D$ matrices in Eq. (\ref{eq:udu_equal_wdw}) are to emphasize the difference from $\Wmat$ and $\Dhat$ in Eq. (\ref{eq:P_wrong_dimensions}).

The upper triangular matrix $ \tilde{\Wmat} $, and the diagonal matrix $ \tilde{\D} $, can be obtained with the execution of a Weighted Modified Gram-Schmidt (WMGS) orthogonalization, given the factorized candidate form: the rows $\w_i$ of $ \Wmat$ and the elements of $\Dhat$ (\cite{simon2006optimal}, Chap. 6). To show this, consider the WMGS procedure that consists of the following recursions:
\begin{align}
	\v_n&= \w_n \  \\\label{eq:v_vectors}
	\v_k &= \w_k - \sum_{j=k+1}^{n}u(k,j)\v_j \quad k = n-1,\dots,1 \ \\ 
	u(k,j) &= \frac{\w_k\Dhat\v_j^T}{\v_j\Dhat\v_j\trans}	\quad j,k = 1,\dots, n \ 
\end{align}
with
\begin{equation}\label{eq:vDv_product}
	\v_k\Dhat\v_j^T  = 0 \quad  \text{when}\  k\neq j \ 
\end{equation}
where the $ \v$'s are row vectors of size $ (n+q)$.
Then, by recognizing that $\w_k$ can be alternatively expressed as, 
\begin{equation}\label{eq:alternative_expression_w}
	\w_k=\v_k +  \sum_{j=k+1}^{n}u(k,j)\v_j \quad k=1, \dots,n \ 
\end{equation}
one can
stack all $n$ vectors $\w$ in matrix form as
\begin{equation}
	\begin{bmatrix}
		\w_1\\
		\vdots\\
		\w_n
	\end{bmatrix}=
	\begin{bmatrix}
		1 & u(1,2) & \dots & u(1,n) \\
		0 & 1 & \ddots& \vdots\\
		\vdots & \ddots & \ddots & u(n-1,n)\\
		0 & \dots & \dots & 1
	\end{bmatrix}
	\begin{bmatrix}
		\v_1\\
		\vdots\\
		\v_n
	\end{bmatrix} \ 
\end{equation}
which indicates that $ \Wmat $ can be constructed via the product of the  $n\times n$ upper triangular matrix $\Umat$ of elements $ u(k,j) $, and the $n \times (n+q)$ orthogonal matrix $ \Vmat $ as
\begin{equation}
	\Wmat = \Umat\Vmat \ 
\end{equation}
With this expression in hand, Eq. (\ref{eq:P_wrong_dimensions}) can now be written as
\begin{equation}\label{eq:covariance_for_comparison}
	\Cov^-=	\Usub{\Dsub}\Usub\trans = \Wmat\Dhat\Wmat\trans = (\Umat\Vmat)\Dhat(\Umat\Vmat)\trans = \Umat[]\ [\Vmat\Dhat\Vmat\trans]\ \Umat\trans  
\end{equation}
Since the product $ \Vmat\Dhat\Vmat\trans $ is constructed according to Eq. (\ref{eq:vDv_product}) and $\Vmat$ is $n\times (n+q)$, the bracketed term $[\Vmat\Dhat\Vmat\trans]$ from the previous equation is diagonal of size $n\times n$. At this point, a term-by-term ``equivalence'' is possible for Eq. (\ref{eq:covariance_for_comparison}) which leads to the sought alternative factors $ \tilde{\Wmat}$ and $ \tilde{\D}$:
\begin{equation}
 \Cov^-=	\Usub{\Dsub}\Usub\trans = \Umat[][] \ [\Vmat\Dhat\Vmat\trans] \ \Umat\trans= \tilde{\Wmat}\tilde{\D}\tilde{\Wmat}\trans\ \end{equation}
 with $ \tilde{\Wmat}=\Umat $ and $ \tilde{\D}=\Vmat\Dhat\Vmat\trans $.
Therefore, the propagated $\Usub$ and $\Dsub$ factors are given by,
\begin{equation}\label{eq:finalU}
	\Usub = \Umat \ 
\end{equation}
and
\begin{equation}\label{eq:finalD}
	\Dsub = \Vmat\Dhat\Vmat\trans \ 
\end{equation}

In summary, using the candidate factors $ \Wmat $ and $ \Dhat $ for the execution of the WMGS routine will provide $ \Umat $ and $ \Vmat $ which are used in Eqs. (\ref{eq:finalU}) and (\ref{eq:finalD}) to obtain the propagated factors $ \Usub $ and $ \Dsub $. Here, the WMGS procedure will be regarded and referenced as a function, such that  WMGS($\Wmat,\Dhat$) indicates its execution with arguments $\Wmat$ and $\Dhat$.
\section{Conventional UD Kalman measurement update}
The reformulation of the standard Kalman filter into the UD Kalman update requires several steps. First, the posterior covariance $\stdvec{P}[][][+]$ is expressed in terms of the prior covariance $\stdvec{P}[][][-]$. To accomplish this, the Kalman gain $\stdvec{K}[][]=\stdvec{P}[][][-]\Hmat\trans(\Hmat\stdvec{P}[][][-]\Hmat\trans+\R)^{-1}$, where $\Hmat_{k} =\partder{\h_{k}}{\x}$, is substituted into the Kalman filter update covariance equation $	\stdvec{P}[][][+]=(\stdvec{I-\stdvec{K}[][]}\Hmat)\stdvec{P}[][][-]$, this is
\begin{align}\label{eq:expanded_cov_update}
	\stdvec{P}[][][+]&=(\stdvec{I-\stdvec{K}[][]}\Hmat)\stdvec{P}[][][-]\nonumber\\
	&=\Cov^--\stdvec{P}[][][-]\Hmat\trans(\Hmat\stdvec{P}[][][-]\Hmat\trans+\R)^{-1}\Hmat\Cov^- \ 
\end{align}
Then, the covariance matrix $\stdvec{P}[][][-]$ is replaced by its UD decomposition i.e., $\stdvec{P}[][][-] = \UDU$. In addition, because in this formulation the measurements are processed sequentially, $\Hmat$ will become a row vector $\Hmat_i$, and $\R$ a scalar, $r_i$ (the $i^{th}$ diagonal element of $\R$). Integrating these changes into Eq. (\ref{eq:expanded_cov_update}) gives
\begin{equation}
	\stdvec{P}[][][+] = \UDU - \UDU\Hmat_i\trans(\Hmat_i\UDU\Hmat_i\trans + r_i)^{-1}\Hmat_i\UDU   
\end{equation}
If we define $ \wsub = \Usub\trans\Hmat_i\trans $ and \hlc{the scalar} $ a_i=(\Hmat_i\UDU\Hmat_i\trans + r_i)^{-1} $, the previous equation can be written as
\begin{equation}\label{eq:pureUD}
	\stdvec{P}[][][+] = \UDU - \UD \ \wsub a_i \ \wsub\trans \ \DU
\end{equation}
and further factorizing the matrices $ \Usub $ and $ \Usub\trans$ leads to
\begin{equation}\label{eq:UD_partial_update_before_udu_operation}
	\stdvec{P}[][][+] = \Usub[\Dsub - \Dsub\wsub a_i\wsub\trans\Dsub]\Usub\trans 
\end{equation}
From Eq. (\ref{eq:UD_partial_update_before_udu_operation}), the term in brackets is symmetric and positive definite and its UD decomposition can be computed. Let the UD decomposition of this term be
\begin{equation}\label{eq:UDU_mathcal}
    \mathcal{U}\mathcal{D}\mathcal{U}\trans=\Dsub - \Dsub\wsub a_i\wsub\trans\Dsub
\end{equation}
such that equation (\ref{eq:UD_partial_update_before_udu_operation}) can now be written as
\begin{equation}
    \stdvec{P}[][][+] = \Usub[\mathcal{U}\mathcal{D}\mathcal{U}\trans]\Usub\trans
\end{equation}
Because the product of two upper triangular matrices with ones in the diagonal is also an upper triangular matrix with ones in the diagonal, we have
\begin{equation}
    \stdvec{P}[][][+] = [\Usub\mathcal{U}]\mathcal{D}[\mathcal{U}\trans\Usub\trans]=\Uplus\Dplus\Uplus\trans
\end{equation}
In other words, the factor $\Usub$ is updated via
\begin{equation}
    \Uplus = \Usub\mathcal{U}
\end{equation}
while the update of the factor $\Dsub$ is equal to $\mathcal{D}$, which results from the UD decomposition of the term $[\Dsub - \Dsub\wsub a_i\wsub\trans\Dsub]$ as per Eq.(\ref{eq:UDU_mathcal}):
\begin{equation}
    \Dplus = \mathcal{D}
\end{equation}

Alternatively to this UD update approach where we perform the UD decomposition of the term $[\Dsub - \Dsub\wsub a_i\wsub\trans\Dsub]$, it is common to find in the literature that the UD Kalman update is done via the \textit{modified} Turner-Agee rank-one update (\hlc{or Carlson update}) \cite{agee1972triangular}. The modified Turner-Agee update is inspired by the same idea of updating the $\Umat$ and $\D$ factors directly to obtain a pair of new $\Umat$ and $\D$, and it also uses the terms present in Eq. (\ref{eq:UD_partial_update_before_udu_operation}). However, it is a more efficient and numerically stable approach. Nonetheless, both the UD decomposition approach shown here, and the execution of the modified Turner-Agee update, will provide the updated factors $\Uplus$ and $\Dplus$. The execution of the modified Turner-Agee can be done implementing the algorithm from Appendix \ref{ap:modified_Agee} given the terms $\Usub$, $\Dsub$,$\w$, and $a_i$.

\textbf{Note: }The \textit{unmodified} or standard Agee-Turner also executes a rank-one update. However, it is numerically less stable if used when the updating term is negative \cite{Grewal2001}. In particular, the negative sign of the updating term $\Dsub - \Dsub\wsub a_i\wsub\trans\Dsub$ in equation \ref{eq:UD_partial_update_before_udu_operation} makes the use of the standard Agee-Turner update less suitable in this case. Otherwise, the standard Agee-Turner update would have been appropriate, e.g. with $\Dsub + \Dsub\wsub a_i\wsub\trans\Dsub$. The implementation of the standard Turner-Agee can be done via \ref{ap:conventional_Agee}.

Table \ref{table:UDPU_filter} summarizes the UD Kalman filter. Note that the UD decomposition needed to initialize the $\Umat$ and $\D$ factors is indicated as udu(), and the modified Agee-Turner as modifiedAgeeTurner(). The UD decomposition mechanization is included in Appendix \ref{ap:udu_decomposition}.

Note that if $m$ measurements are available for processing, the UD Kalman update step needs to be executed $m$ times, one execution per measurement. It is important to mention that the UD update step from Table \ref{table:UDPU_filter} assumes uncorrelated measurements, i.e., a diagonal $\R$, and if correlated measurements are present, i.e., if $\R$ is not diagonal, the practitioner can execute a decorrelation procedure prior processing the measurements. The mechanization included in Appendix \ref{ap:decorrelation} can be used to decorrelate the available measurements (diagonalize $\R$) so that the UD update from Table \ref{table:UDPU_filter} can still be used.

\begin{table}[h!]
	\caption{U-D Kalman filter update with sequential uncorrelated measurement processing}\label{table:UDPU_filter}
	\centering
	\setlength{\extrarowheight}{8pt}
	\begin{tabular}{ |c|c| } 
		\hline
		\textbf{Model} & $\begin{aligned} &\x_{k}=\f_{k-1}(\x_{k-1},\u_{k-1},\w_{k-1})\\ &\yktilde=\hkofxk+\vk\\
			&\wk \sim \mathcal{N} (\zerovec,\Qmat_k)\\
			&\vk \sim \mathcal{N}(\zerovec,\R_k)\end{aligned}$\\
		\hline
		\textbf{Initialize} & $\begin{aligned} 
			&\xpost_0=\x_0 \\
			&[\Uplus_0,\Dplus_0]= \text{udu}(\stdvec{P}[][0][+])\ \text{[Appendix \ \ref{ap:udu_decomposition}]}\\
		\end{aligned}$\\
		\hline
		\textbf{UD Propagation} & $\begin{aligned} &\xprior_{k}=\f_{k-1}(\xpost_{k-1},\u_{k-1})\\
			&\Wmat_{k-1} = \begin{bmatrix}
				\Fmat_{k-1}\Uplus_{k-1} && \Gmat_{k-1}
			\end{bmatrix}\ ;
			\Dhat_{k-1} = \begin{bmatrix}\Dplus_{k-1} && \zerovec\\
				\zerovec && \Qmat_{k-1}
			\end{bmatrix}\\
			&[\Usub_{k},\Dsub_{k}]=\text{WMGS}(\Wmat_{k-1},\Dhat_{k-1}) \ \text{[Section \ref{subsec:overview_UD}}] \\
		\end{aligned}$\\
		\hline
		\textbf{UD update} & 
		$\begin{aligned}
			[\Uplus_k,\Dplus_k,\Kmat_k] &=\text{ modifiedAgeeTurner}(\Usub_k, \Dsub_k, r_{i_k},  \Hmat_{i_k})\ \text{[Appendix \ \ref{ap:modified_Agee}]} \\
			\xpost_k&=\xprior_{k}+\stdvec{K}[][k](\ym_k-\h(\xhat_k^-))\\
		\end{aligned}$\\
		\hline
		\end{tabular}
\end{table}

\section*{Appendix}

\subsection{UD decomposition}\label{ap:udu_decomposition}
Given a symmetric, positive-definite, $m\times m$ matrix $\stdvec{M}$, its UD decomposition can be obtained via the following mechanization. 
\begin{flalign}
\textbf{for}&\ j=m:-1:1&&\\
    &\textbf{for}\ i=j:-1:1&&\\
        & \quad \sigma=M(i,j)&&\\
        & \quad \textbf{for}\ k=j+1:m&&\\
                &\qquad \sigma=\sigma-U(i,k)D(k,k)U(j,k)&&\\
        & \quad \textbf{end for}&&\\
        &\quad \textbf{if}(i==j)&&\\
        &\qquad D(j,j)=\sigma&&\\
        &\qquad U(j,j)=1&&\\
        &\quad\textbf{else}&&\\
        &\qquad U(i,j)=\sigma/D(j,j)&&\\
        &\quad\textbf{end if}&&\\
        &\ \ \textbf{end for}&&\\
        &\textbf{end for}\\
   \end{flalign}
   
The $\Umat$ and $\D$ matrices constructed via this mechanization satisfy $\stdvec{M}=\Umat\D\Umat\trans$. Note that in Table \ref{table:UDPU_filter}, this algorithm is referred as the \textit{function} \text{udu()}.

\subsection{Modified Agee-Turner rank-one update}\label{ap:modified_Agee}
The UD Kalman filter uses the following algorithm to update the UD factors. Providing $\Usub$, $\Dsub$, $r_i$, $\Hmat_i$, the algorithm returns $\Uplus$, the elements of $\Dplus_j$ as [$d_1 \ d_2 \ \hdots d_n$], and the Kalman gain $\Kmat$. This approach is taken from \cite{Gibbs2011a}.
  \begin{flalign}
    &\w = \Usub\trans\Hmat_i \ \text{where} \ \w = [w_1\ w_2 \ \hdots \ w_n]\trans&&\\
    &\vchar = \Dsub\trans\w \ \text{where} \ \vchar=[\barbelow{d}_1 w_1\ \barbelow{d}_2 w_2 \ \hdots \ \barbelow{d}_n w_n]\trans \\
    &\Kmat_1 = [v_1\ 0\ \hdots \ 0]\trans \\
    &\alpha_1 = r_i+v_1w_1 \\
    &d_1 = (r_i\barbelow{d}_1)/\alpha_1\\
    &\textbf{for}\ j=2 \hdots \ n &&\\
    &\ \ \ \ \alpha_j = \alpha_{j-1}+v_jw_j\\
    &\ \ \ \ d_j = d_j\alpha_{j-1}/\alpha_j\\
    &\ \ \ \ \lambda_j = -w_j/\alpha_{j-1}\\
    &\ \ \ \ \Uplus_j = \Usub_j + \lambda_j\Kmat_{j-1}\\
    &\ \ \ \ \Kmat_j = \Kmat_{j-1}+v_j\Usub_j\\
    &\textbf{end for}\\
    &\text{Finally compute}\nonumber\\
    &\Kmat = \Kmat_{n}/\alpha&&
  \end{flalign}
In this algorithm, $\Uplus_j$, $\Usub_j$, and $\Kmat_{j}$, refer to the $j^{th}$ column of the corresponding matrix. Note that in Table \ref{table:UDPU_filter}, this algorithm is referred as the \textit{function} \text{modifiedAgeeTurner()}.

\subsection{The Agee-Turner rank-one update}\label{ap:conventional_Agee}
This algorithm provides a mechanization to update the $n\times n$ factors $\Umat$ and $\D$ with the information contained in $c\stdvec{a}\stdvec{a}\trans$ according to 
\begin{equation}
    \overset{+}{\stdvec{U}}\overset{+}{\stdvec{D}}\overset{+}{\stdvec{U}}\trans=\Umat\D\Umat\trans + c\stdvec{a}\stdvec{a}\trans .
\end{equation}
Where c is a positive scalar and $\stdvec{a}$ is a column vector of size $n$. This algorithm returns the updated factors $\overset{+}{\stdvec{U}} $ and $\overset{+}{\stdvec{D}}$ given $\Umat, \D,c$, and $\stdvec{a}$. The algorithm is executed as follows:

\begin{flalign}
&C_n = c&&\\
&\textbf{for}\ j=n,n-1,\hdots,2\\
&\ \ \ \ \Uplus_{jj}=1\\
&\ \ \ \ \Dplus_{jj} = \D_{jj}+C_ja_j^2\\
&\ \ \ \ \textbf{for}\ k=1, \hdots, j-1\\
&\ \ \ \ \ \ \ a_k \coloneqq a_k-a_j\Umat_{k,j}\\
&\ \ \ \ \ \ \ \Uplus_{kj}=\Umat_{kj}+c_j a_j a_k/\Dplus_{jj}\\
&\ \ \ \ \ \textbf{end for}\\
&\ \ \ C_{j-1} = C_j\D_{jj}/\Dplus_{jj}\\
&\textbf{end for}\\
&\Dplus_{11}=\D_{11}+C_1a_1^2
\end{flalign}
Note that all of the sub-indices are used to access the matrix elements.

\subsection{Measurement decorrelation using UD factors}\label{ap:decorrelation}
	Let the UD decomposition of a non-diagonal measurement noise covariance, $ \R_c $, be
	\begin{equation}
	\R_c = \Ur\Dr\Ur\trans \ 
	\end{equation}
	and consider the measurement equation $\y = \Hmat\x+\v$, being transformed by $\Uinv$ as
	\begin{equation}	
	\stdvec{z} = \Uinv\y = \Uinv\Hmat\x+\Uinv\v \ 
	\end{equation}
	The residual (measurement minus expected measurement) can then be calculated as
	\begin{equation}
	\stdvec{e}=\stdvec{z}-\hat{\stdvec{z}}=\Uinv\Hmat(\stdvec{x}-\hat{\stdvec{x}})+\Uinv\v \ 
	\end{equation}
	and its corresponding covariance as
	\begin{equation}
	\expect[\stdvec{e}\stdvec{e}\trans]=\expect[(\Uinv\Hmat(\stdvec{x}-\hat{\stdvec{x}})+\Uinv\v)(\Uinv\Hmat(\stdvec{x}-\hat{\stdvec{x}})+\Uinv\v)\trans ] \ 
	\end{equation}
	\begin{equation}
	\expect[\stdvec{e}\stdvec{e}\trans]=\Uinv\Hmat\expect[(\stdvec{x}-\hat{\stdvec{x}})(\stdvec{x}-\hat{\stdvec{x}})]\trans\Hmat\trans\Ur^{-T} +\Uinv\expect[\v\v]\trans\Ur^{-T} \ 
	\end{equation}
	\begin{equation}
	\expect[\stdvec{e}\stdvec{e}\trans]=\Uinv\Hmat\Cov\trans\Hmat\trans\Ur^{-T} +\Uinv\R_c\Ur^{-T} \ 
	\end{equation}
	 assuming zero mean Gaussian white noise.
	But since
	\begin{equation}
	\Uinv\R_c  \Ur^{-T}=\Dr \ 
	\end{equation} 
	and by naming $\Hmat_z=\Uinv\Hmat$, this results in
	\begin{equation}
	\expect[\stdvec{e}\stdvec{e}\trans] = \Hmat_z\Cov\Hmat_z\trans+\Dr \ 
	\end{equation}
	That is, the transformed equation now uses a diagonal measurement noise covariance. In order to properly use the transformed, now uncorrelated measurements, the filter simply uses $ \Uinv\Hmat $ instead of $ \Hmat $, and instead of directly computing the measurement residual with $ (\y-\Hmat\x)$ or $ (\y-\h(\xprior))$ , we now use $\Uinv(\y-\Hmat\x)$. Rather than using the correlated measurement noise $ \R_c $, the filter now will use $ \Dr $.

	\section*{Funding Sources}
 This work was supported under Air Force contract FA8651-20-F-1025.


	\bibliography{final_udu.bib}

\end{document}